\documentclass[%
 reprint,
 amsmath,
 aps,
]{revtex4-2}

\usepackage{graphicx}
\usepackage{dcolumn}
\usepackage{bm}

\usepackage{amsmath}
\usepackage[colorlinks,
 linkcolor=blue,
 anchorcolor=blue, 
 citecolor=blue,  
]{hyperref}

\begin{document}

\title{Strengthening practical continuous-variable quantum key distribution systems against measurement angular error}

\author{Tao Shen$^{{1},{2}}$}
\author{Yundi Huang$^{1}$}
\author{Xiangyu Wang$^{1}$}
\thanks{xywang@bupt.edu.cn}
\author{Huiping Tian$^{{1},{2}}$}
\author{Ziyang Chen$^3$}
\thanks{chenziyang@pku.edu.cn}
\author{Song Yu$^{1}$}

\affiliation{$^1$State Key Laboratory of Information Photonics and Optical Communications, Beijing University of Posts and Telecommunications, Beijing 100876, China}
\affiliation{$^2$School of Information and Communication Engineering, Beijing University of Posts and Telecommunications, Beijing 100876, China}
\affiliation{$^3$State Key Laboratory of Advanced Optical Communication Systems and Networks, Department of Electronics, and Center for Quantum Information Technology, Peking University, Beijing 100871, China}

\date{\today}

\begin{abstract}

The optical phase shifter that constantly rotates the local oscillator phase is a necessity in continuous-variable quantum key distribution systems with heterodyne detection.
In previous experimental implementations, the optical phase shifter is generally regarded as an ideal passive optical device that perfectly rotates the phase of the electromagnetic wave of $90^\circ$.
However, the optical phase shifter in practice introduces imperfections, mainly the measurement angular error, which inevitably deteriorates the security of the practical systems.
Here, we will give a concrete interpretation of measurement angular error in practical systems and the corresponding entanglement-based description.
Subsequently, from the parameter estimation, we deduce the overestimated excess noise and the underestimated transmittance, which lead to a reduction in the final secret key rate.
Simultaneously, we propose an estimation method of the measurement angular error.
Next, the practical security analysis is provided in detail, and the effect of the measurement angular error and its corresponding compensation scheme are demonstrated.
We conclude that measurement angular error severely degrades the security, but the proposed calibration and compensation method can significantly help improve the performance of the practical CV-QKD systems.

\end{abstract}

\maketitle


\section{\label{sec:level1}Introduction }

Continuous-variable quantum key distribution (CV-QKD) provides a way for two remote participants called Alice and Bob to establish
symmetric keys through an unsafe channel \cite{weedbrook2012gaussian,grosshans2003quantum}. In CV-QKD, the information is encoded in the quadratures of the optical field and decoded by coherent detection.
The physical implementation of CV-QKD using Gaussian modulated
coherent state (GMCS) is based on mature optical communication techniques with high reliability and low cost. Thus, CV-QKD has attracted much attention in recent years.
Theoretical research of GMCS CV-QKD has made great progress \cite{furrer2014reverse,leverrier2015composable,leverrier2017security}, and such significant progress accelerates experimental progress  \cite{jouguet2013experimental,huang2016field,huang2016long,wang2017experimental,zhang2019continuous,zhang2020long}. As well, an overview of practical CV-QKD systems using telecom components is proposed \cite{guo2021toward}.

The security analysis of CV-QKD contains theoretical security analysis, practical security analysis and composable security analysis under the universal framework.
Although the theoretical security and composable security of CV-QKD protocol based on GMCS have been proved based on many assumptions \cite{leverrier2013security,leverrier2015composable,leverrier2017security}, the practical security problems caused by imperfections of experimental devices remain unsolved absolutely. Nevertheless, the non-ideal elements of the system are so numerous that they are often ignored.
What is worse, the attacks controlled by Eve against the system never end. Thus, researchers put forward continuous-variable measurement-device-independent (CV-MDI) or continuous-variable one-sided-device-independent (CV-1sDI) protocol to resist those attacks involving the imperfections of devices  \cite{walk2016experimental,gehring2015implementation,pirandola2015high}. However, the experimental implementation of CV-MDI or CV-1sDI protocol is somewhat tricky.
As a result, using proven CV-QKD protocols and dealing with the imperfections of experimental devices within these protocols is the most efficient method \cite{huang2014quantum,jouguet2013preventing,ma2013local,zhao2018polarization}.
These imperfections may not be eradicated absolutely, but there are corresponding methods to mitigate the impacts of these non-ideal factors. For instance, imperfect Gaussian modulation due to the jitter of the half-wave voltage of the intensity modulator and the phase modulator can be solved by a unique calibration method \cite{liu2017imperfect}. One-time calibration model solves imperfect monitoring at light source \cite{chu2021practical}. Attacks from the poor linearity of the homodyne detector and imperfect beam splitter are defended by adding additional monitoring devices \cite{qin2016quantum,huang2013quantum}.

In the GMCS CV-QKD protocol, the receiver usually adopts homodyne or heterodyne detection. As for the homodyne detection scheme, only one quadrature component $X$ or $P$ needs to be measured, so basis choice is an indispensable step, and imperfect basis choice will cause security problems \cite{liu2020imperfect}. In the heterodyne detection scheme, both quadrature components $X$ and $P$ need to be measured, which is similar to phase diversity reception in classical optical communication \cite{painchaud2009performance,zhang2012towards}.
In the CV-QKD scheme using heterodyne detection, we usually choose the optical phase shifter to help measure the $P$ quadrature. The optical phase is demanded to rotate the phase by $90^\circ$ constantly.
Here we restricted ourselves to optical fibre systems for simplicity. That is to say, the optical phase shifter in this paper is an optical fibre phase shifter, short for phase shifter.
However, under the action of external force, the fibre is stretched or compressed within the elastic deformation range, and parameters such as the fibre change's geometrical size and refractive index change, thus causing the phase change of the transmitted signal in the fibre. Therefore, the phase shifter is somewhat susceptible to environmental changes and can hardly shift the phase by $90^\circ$ exactly, resulting in a deviation of the true value from $90^\circ$, which is defined as the measurement angular error in this paper.

We start with expressions of the canonical quadratures of the received state and give the corresponding theoretical model of CV-QKD using heterodyne detection.
The security analysis and simulation show that the secret key rates would be decreased in the presence of measurement angular error.
Furthermore, we propose a method that can calibrate the measurement angular error. Thus it can close the potential security loopholes.

The paper is organized as follows. In Sec. \ref{sec:level2}, we provide a practical implementation analysis and entanglement-based (EB) description of measurement angular error, and then a compensation method is provided. In Sec. \ref{sec:level3}, we discuss the practical security of measurement angular error through two aspects; one is the estimation of quantum channel parameters, the rest is the detailed calculation of secret
key rates. Then in Sec. \ref{sec:IV}, numerical simulation results and analysis are provided. In Sec. \ref{sec:V}, we give a conclusion and discuss the importance of compensation on measurement angular error.

\section{\label{sec:level2}Measurement angular error in practical CV-QKD }

\subsection{\label{sec:level2_1}Practical implementation analysis of measurement angular error}

In the GMCS CV-QKD protocol, Alice randomly generates two groups of Gaussian random numbers which corresponds to $X$ and $P$ quadratures. They have the same modulation variance ${V_A}$ in shot-noise units (SNUs). Gaussian modulation consists of intensity modulation and phase modulation. Intensity obeys a Rayleigh distribution, while phase obeys a uniform distribution.

\begin{equation}\label{eq.1}
\begin{array}{l}
X = {A_{sig}}\cos {\phi _{sig}},\\
P = {A_{sig}}\sin {\phi _{sig}},
\end{array}
\end{equation}
where ${A_{sig}}$ and ${\phi _{sig}}$ are the modulation information loaded on the intensity modulator and the phase modulator respectively.

\begin{figure}[t]
\centering
\includegraphics[width=11cm,height=6.5cm]{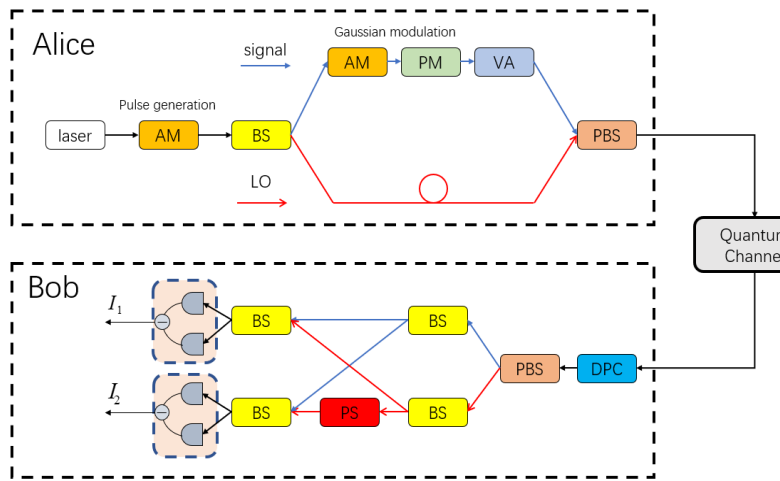}
\caption{Schematics layout of the heterodyne detection GMCS protocol. AM: amplitude modulator; PM: phase modulator; BS: beam splitter; PBS: polarization beam splitter; VA: variable attenuator; DPC: dynamic polarization controller; PS: phase shifter.}\label{fig_1}
\end{figure}

The practical implementation scheme of GMCS protocol with heterodyne detection can be seen in Fig. \ref{fig_1}. To prepare the quantum state, Alice generates continuous light using a laser source. The first AM is just for pulse generation, and then the pulse is divided into two splits with a beam splitter (BS); one travels through the signal path, the other travels through the local oscillator (LO) path. After Gaussian modulation and proper attenuation, the quantum signal is polarized multiplexed with the LO using a polarization beam splitter (PBS). The prepared signal are then transmitted to Bob through a noisy quantum channel. At the receiver side, the signal is demultiplexed with a PBS and a dynamic polarization controller (DPC). The heterodyne detection scheme comprises the optical phase shifter and two balanced homodyne detectors (BHDs); one measures $X$ quadrature while the other measures $P$ quadrature. In the CV-QKD scheme using heterodyne detection , the optical phase shifter needs to be set to $90^\circ $ to measure the quadrature component $P$. But the actual shifting phase is $ \varphi _{PS} =\frac{\pi }{2} - \theta $ ($\theta$ is a small deviation value) due to the non-ideal external factors. Other irrelevant devices and operations are assumed to be ideal for simplicity. The electric field expression of signal and LO before detection can be expressed by \cite{corvaja2017phase}
\begin{align}\label{eq.2}
{E_{sig1}} &= {E_{sig2}} = {A_{sig}}\cos \left( {2\pi {f_s}t + {\phi _{sig}} + {\varphi _0} + {\varphi _{channel\_sig}}} \right)\notag,\\
{E_{L{O_1}}} &= {A_{LO}}\cos \left( {2\pi {f_L}t + {\varphi _0} + {\varphi _{channel\_LO}}} \right),\\
{E_{L{O_2}}} &= {A_{LO}}\cos \left( {2\pi {f_L}t + {\varphi _0} + {\varphi _{channel\_LO}} + {\varphi _{PS}}} \right)\notag,
\end{align}
where $E$ represents the electric field intensity, ${f_s}$ and ${f_L}$ are the center frequency of the quantum signal and the LO, ${\varphi _0}$ is the initial phase of the laser and
${\varphi _{channel}}$ represents the phase during quantum channel. Thus, the generated photocurrents of the first and second homodyne detectors can be expressed as ${I_1}$ and ${I_2}$ \cite{wang2020high}, respectively
\begin{align}\label{eq.3}
{I_1} &= {R_1}{\eta _1}\left( {{{\left| {{E_{sig1}} + {E_{L{O_1}}}} \right|}^2} - {{\left| {{E_{sig1}} - {E_{L{O_1}}}} \right|}^2}} \right)\notag\\
 &= {R_1}{\eta _1}{A_{sig}}{A_{LO}}\cos \left( {2\pi \Delta ft + {\phi _{sig}} + \Delta {\varphi _{channel}}} \right),\\
{I_2} &= {R_2}{\eta _2}\left( {{{\left| {{E_{sig2}} + {E_{L{O_2}}}} \right|}^2} - {{\left| {{E_{sig2}} - {E_{L{O_2}}}} \right|}^2}} \right)\notag\\
 &= {R_2}{\eta _2}{A_{sig}}{A_{LO}}\cos \left( {2\pi \Delta ft + {\phi _{sig}} + \Delta {\varphi _{channel}} + {\varphi _{PS}}} \right)\notag,
\end{align}
where ${R_i}$ and ${\eta _i}$ denote the responsiveness and quantum efficiency of two balanced homodyne detectors ($i=1,2$). Because the quantum signal and the LO are from the same laser and their optical path is almost the same, so we consider $\Delta f=0$. Meantime, we assume the shifting phase in a quantum channel of the quantum signal is the same as that of the LO, so $\Delta {\varphi _{channel}}=\varphi _{channel\_sig}-\varphi _{channel\_LO}=0$.
By taking Eq. (\ref{eq.1}) into Eq. (\ref{eq.3}),
the photocurrents can be further be written as
\begin{equation}\label{eq.4}
\begin{array}{l}
{I_1} = {R_1}{\eta _1}{A_{LO}}{X_1},\\
{I_2} = {R_2}{\eta _2}{A_{LO}}(\sin \theta  \cdot {X_2} - \cos \theta  \cdot {P_2}),
\end{array}
\end{equation}
where ${X_i}$ and ${P_i}$ are measurement results on the upper branch ($i=1$) or lower branch ($i=2$). Since the phase shifter  can only change the phase and does not influence the intensity of the SNUs, so the coefficient
${R_i}{\eta _i}{A_{LO}}$ of $I_1$ and $I_2$ would be normalized by SNUs directly. Thus, the normalized measurement results $I_1'$ and $I_2'$ can be expressed as
\begin{equation}\label{eq.5}
\begin{array}{l}
{I_1}' = {X_1},\\
{I_2}' = \sin \theta  \cdot {X_2} - \cos \theta  \cdot {P_2}.
\end{array}
\end{equation}
In doing so, the influence of the imperfect phase shifter on the final heterodyne detection can be seen more intuitively in the form of photocurrent.

\subsection{\label{sec:level2_2}Entanglement-Based description of measurement angular error}

\begin{figure}[tb]
\centering
\includegraphics[width=10cm,height=4cm]{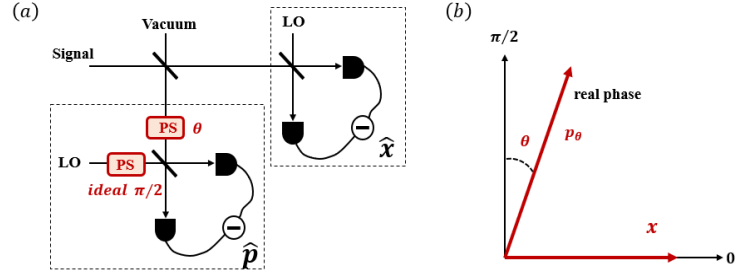}
\caption{Model of real heterodyne detector with measurement angular error in CV-QKD. PS: phase shift.
(a) Model of non-ideal $P$ quadrature measurement. (b) Non-ideal
measurement in phase space.}\label{fig_2}
\end{figure}

In this subsection, we mainly discuss the EB description of the measurement angular error in the CV-QKD scheme using heterodyne detection. In Sec. \ref{sec:level2_1}, we have shown that the phase of LO is rotated by
$ \varphi _{PS} =\frac{\pi }{2} - \theta $ in practical heterodyne detection scheme caused by imperfect phase shifter. In the new model, it can be interpreted that the phase of LO is rotated $90^\circ$ ideally while the quantum signal following a phase shifts $\theta$, as shown in Fig. \ref{fig_2}(a). Equivalently, this means we can first rotate the state to be measured by $\theta$ and then perform the ideal measurement to obtain the $P$ quadrature in phase space, as illustrated in Fig. \ref{fig_2}(b). The quadrature transformation of the phase shift operator is given in term of the symplectic matrix ${S_{PS}}$, which reads
\begin{equation}\label{eq.6}
{S_{PS}} = \left( {\begin{array}{*{20}{c}}
{\cos \theta }&{\sin \theta }\\
{ - \sin \theta }&{\cos \theta }
\end{array}} \right).
\end{equation}

\begin{figure}[b]
\centering
\includegraphics[width=10cm,height=4cm]{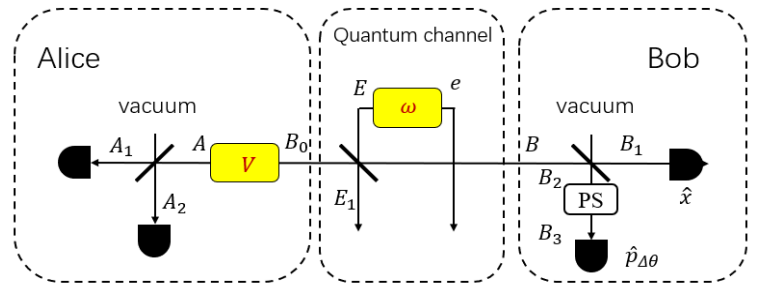}
\caption{The EB description of measurement angular error in CV-QKD scheme using heterodyne detection.}\label{fig_3}
\end{figure}

Now we present the complete EB model of measurement angular error in the CV-QKD scheme using heterodyne detection. The coherent state preparation by Alice is modelled by a heterodyne measurement of one half of a two-mode squeezed vacuum (EPR) state of variance $V$. The other half of the EPR state is sent to Bob through the quantum channel controlled by Eve. At the receiver, Bob performs the heterodyne detection on the mode $B$ and uses a 50:50 BS to split mode $B$ into two different modes, namely, $B_1$ and $B_2$, and each of them is measured using homodyne detection. Among them, a phase shift operation is performed while measuring $P$ quadrature, as shown in Fig. \ref{fig_3}. The heterodyne detection results can be expressed as
\begin{align}
\label{eq.8}{x_{{B_1}}} &= \frac{1}{{\sqrt 2 }}{x_B} + \frac{1}{{\sqrt 2 }}{x_v},\\
\label{eq.9}{p_{{B_3}}} &=  - \sin \theta  \cdot {x_{{B_2}}} + \cos \theta  \cdot {p_{{B_2}}},
\end{align}
where ${x_{{B_2}}} =- \frac{1}{{\sqrt 2 }}{x_B} + \frac{1}{{\sqrt 2 }}{x_v}$, ${p_{{B_2}}} =- \frac{1}{{\sqrt 2 }}{p_B} + \frac{1}{{\sqrt 2 }}{p_v}$, which could be calculated from quadrature transformation ${S_{BS}}$ in Sec. \ref{sec:level3}.

It can then be observed from Eq. (\ref{eq.5}) and Eq. (\ref{eq.9}) that the EB description above is equivalent to the practical implementation of measurement angular error in the CV-QKD scheme using heterodyne detection.

\subsection{\label{sec:level2_3}Estimation of measurement angular error}
In this subsection, we propose a method to estimate the measurement angular error in the GMCS CV-QKD scheme using heterodyne detection. From Sec. \ref{sec:level2_2}, we have illustrated that the EB description is equivalent to the PM description in this scheme.
Therefore, the results in the PM model can be substituted into the corresponding EB model.
In the meanwhile, according to the EB model, there are relationships between the various modes, from which the derivation of the measurement angular error can be found.
Hence, according to Eq. (\ref{eq.8}) and Eq. (\ref{eq.9}), we can acquire the concrete formulation about covariance and variance, which can be measured directly from the practical experiment
\begin{gather}
\label{eq.20}\left\langle {{p_{{B_3}}},{x_{{B_1}}}} \right\rangle  =  - \sin \theta  \cdot ( - \frac{{{V_B}}}{2} + \frac{1}{2}),\\
\label{eq.10}{V_{{B_1}}} = \frac{1}{2}\left( {{V_B} + 1} \right),
\end{gather}
where $\left\langle {{p_{{B_3}}},{x_{{B_1}}}} \right\rangle$ is the covariance between ${x_{{B_1}}}$ and ${p_{{B_3}}}$, and
$V_{B_1}$ is the variance of ${x_{{B_1}}}$. The values above are all normalized by SNUs.
Then, the mean value of the error angle can be simplified as
\begin{equation}\label{eq.11}
\theta  = \arcsin \left( {\frac{{\left\langle {{p_{{B_3}}},{x_{{B_1}}}} \right\rangle }}{{{V_{{B_1}}} - 1}}} \right).
\end{equation}

In order to figure out $p_{B_2}$, the ideal measurement result of $P$ quardrature, and the precise calibration of $\theta$ is essential as well as $x_{B_2}$.
Since the BS will introduce a vacuum state, there exists an uncertainty from the vacuum fluctuation.
However, the deviation is not dominant in the measurement result; thus, it allows us to assume $x_{{B_1}}\approx{x_{{B_2}}}$,
then $p_{B_2}$ can be calculated, and the measurement angular error can be compensated through post-processing as in Eq. (\ref{eq.9}).

\section{\label{sec:level3}practical security analysis of measurement angular error}

Collective attack is regarded as an optimal attack among Gaussian-modulation coherent detection schemes \cite{leverrier2015composable,leverrier2017security}.
Hence, the analysis about practical security under collective attack is worthy of discussing.
Thus, the security analysis against the collective attack of the CV-QKD protocol with heterdyne detection is provided.

In the system with imperfect phase shifter, the parameters required for security analysis are needed to be obtained through parameter estimation in the PM model.
Therefore, we briefly describe the process of parameter estimation at first.
For a general linear channel, the correlation of data between Alice and Bob is given by
\begin{equation}\label{eq.12}
y = tx + z,
\end{equation}
where $x$ is the Gaussian modulation quantum signal with the modulation variance $V_A$, $y$ is the received quantum signal through a Gaussian channel and balanced homodyne detectors (BHDs) with total transmittance $t$. $z$ contains excess noise from quantum channel and electrical noise from BHDs.

Thus, in ideal-measurement scenario, measurement results after the heterodyne detection can be written as
\begin{equation}\label{eq.13}
\left\{ {\begin{array}{*{20}{c}}
{{x_B} = {t_x}{x_A} + {z_x}},\\
{{p_B} = {t_p}{p_A} + {z_p}},
\end{array}} \right.
\end{equation}
$\sqrt {{\eta _1}{T_x}} $
where ${t_x} = \sqrt {{\eta _1}{T_x}}$, ${t_p}=\sqrt {{\eta _2}{T_p}}$. The quantum channel parameters of transmittance $T$ and excess noise $\varepsilon $ are relevant to those values through following equations
\begin{gather}
\label{eq.14}\left\langle {{x_A},{x_B}} \right\rangle  = \sqrt {{\eta_1} {T_x}}  \cdot {V_A} ,\left\langle {{p_A},{p_B}} \right\rangle = \sqrt {{\eta_2} {T_p}}  \cdot {V_A}   ,\\
\label{eq.15}{\left\langle {{x_B}^2} \right\rangle  = {\eta_1} {T_x} \cdot {V_A} + {\eta_1} {T_x} \cdot {\varepsilon_1}  + 1 + {v_{el}}_1},\\
\label{eq.16}{\left\langle {{p_B}^2} \right\rangle  = {\eta_2} {T_p} \cdot {V_A} + {\eta_2} {T_p} \cdot {\varepsilon_2}  + 1 + {v_{el}}_2},
\end{gather}
where ${V_A} = \left\langle {{x_A}^2} \right\rangle =\left\langle {{p_A}^2} \right\rangle $, $\eta_i $ and ${{v_{el}}_i}$ are performance parameters of the the BHDs, we assume BHDs in our model are ideal, so the detection efficiency ${\eta_1} = {\eta_2}=1 $, electronic noise ${v_{el}}_1 = {v_{el}}_2=0$.
Noted that all the parameters above have been normalized by SNUs.
The secret key rate under collective attack is derived
from the covariance matrix \cite{weedbrook2012gaussian}
\begin{equation}\label{eq.50}
{\gamma _{A{B_1}{B_3}}} = \left( {\begin{array}{*{20}{c}}
  {{V_A}}&{{C_{A{B_1}}}}&{{C_{A{B_3}}}} \\
  {C_{A{B_1}}^T}&{{V_{{B_1}}}}&{{C_{{B_1}{B_3}}}} \\
  {C_{A{B_3}}^T}&{C_{{B_1}{B_3}}^T}&{{V_{{B_3}}}}
\end{array}} \right).
\end{equation}
$V$ and $C$ in this matrix represent the variance of and correlation of the mode.
The transmittance $T$ and excess noise $\varepsilon $ in the covariance matrix can then be derived with the data of Alice and the heterodyne detection results of Bob
\begin{align}
\label{eq.17}
\left\{ {\begin{array}{*{20}{c}}
  {{T_{x(p)}} = \frac{{{{\left\langle {x{{(p)}_A},x{{(p)}_B}} \right\rangle }^2}}}{{{{\left\langle {x{{(p)}_A}^2} \right\rangle }^2}}},} \\
  {{\varepsilon _{x(p)}} = \frac{{\left\langle {x{{(p)}_B}^2} \right\rangle  - 1}}{{{T_{x(p)}}}} - \left\langle {x{{(p)}_A}^2} \right\rangle .}
\end{array}} \right.
\end{align}

In the case of measurement angular error, one should replace
${p_B}$ with ${p_B}'$ in Eq. (\ref{eq.9}), $T_p$ and $\varepsilon _p$ in Eq. (\ref{eq.17}) should subsequently be rewritten as
\begin{equation}\label{eq.19}
\left\{ {\begin{array}{*{20}{c}}
{{T_p}' = {{\cos }^2}\theta  \cdot {T_p}},\\
{{\varepsilon _p}' = \frac{{{\varepsilon _p}}}{{{{\cos }^2}\theta }} + {{\tan }^2}\theta  \cdot \left\langle {{x_A}^2} \right\rangle}.
\end{array}} \right.
\end{equation}

It is obvious that the estimated ${T_p}'<{T_p}$  and ${\varepsilon_p}'>{\varepsilon_p}$ if the measurement angular error is above zero. Thus, covariance matrix becomes
\begin{equation}\label{eq.51}
{\gamma _{A{B_1}{B_3}}} = \left( {\begin{array}{*{20}{c}}
  {{V_A}}&{{C_{A{B_1}}}}&{{C_{A{B_3}}}'} \\
  {C_{A{B_1}}^T}&{{V_{{B_1}}}}&{{C_{{B_1}{B_3}}}'} \\
  {C_{A{B_3}}^T}'&{C_{{B_1}{B_3}}^T}'&{{V_{{B_3}}}'}
\end{array}} \right).
\end{equation}

By comparing the above two covariance matrices Eq.(\ref{eq.50}) and Eq.(\ref{eq.51}), the different elements is ${C_{A{B_3}}},{C_{{B_1}{B_3}}}$ and ${V_{{B_3}}}$ which contains
transmittance ${T_p}'$ and excess noise ${\varepsilon_p}'$ which are the decisive parameters for the secret key rate of CV-QKD.
The detailed calculation of secret key rate is shown in the Appendix \ref{sec:VI}.

\section{\label{sec:IV}Simulation and results}

\begin{figure}[b]
\centering
\includegraphics[width=8.6cm,height=6cm]{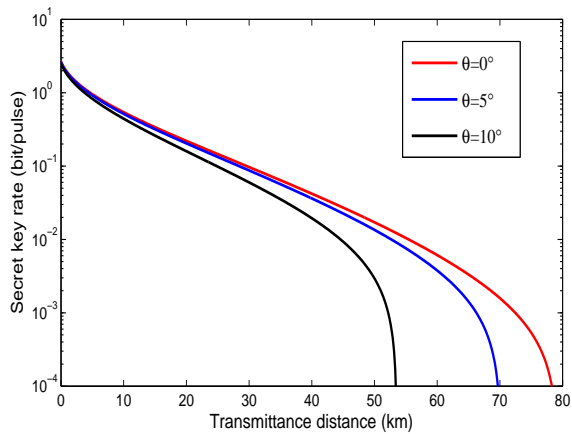}
\caption{The secret key rate versus the transmission
distance with different measurement angular error. From left to right, the curves correspond to secret key rate for
$\theta  = {0^ \circ }$, $\theta  = {5^ \circ }$ and $\theta  = {10^ \circ }$. The other parameters of systems: modulation variance is $V=20$, excess noise is $\varepsilon  = 0.01$, reconciliation efficiency is $\beta  = 0.95$.}\label{fig_4}
\end{figure}

In this section, we illustrate the simulation results about the effects of measurement angular error given the security analysis in Sec. \ref{sec:level3}. First of all, what needs to be simulated is the influence of measurement angular error on the system, among which the most important one is its influence on the secret key rate. Thus, the secret key rate as a function of transmittance distance with different measurement angular errors is calculated, as depicted in Fig. \ref{fig_4}.
Three colors of lines, red, blue, and black correspond to different angles $\theta  = {0^ \circ }$ (after compensation), $\theta  = {5^ \circ }$ and $\theta  = {10^ \circ }$, in this simulation, channel excess noise is set to be $\varepsilon  = 0.01$, and the reconciliation efficiency
$\beta  = 0.95$ \cite{wang2017efficient}. From the Fig. \ref{fig_4}, it can be seen that the secret key rate with measurement angular error is lower than that in the ideal scenario. Furthermore, the secret key rate decreases sharply with the increase of measurement angular error.

\begin{figure}[t]
\centering
\includegraphics[width=8.7cm,height=6cm]{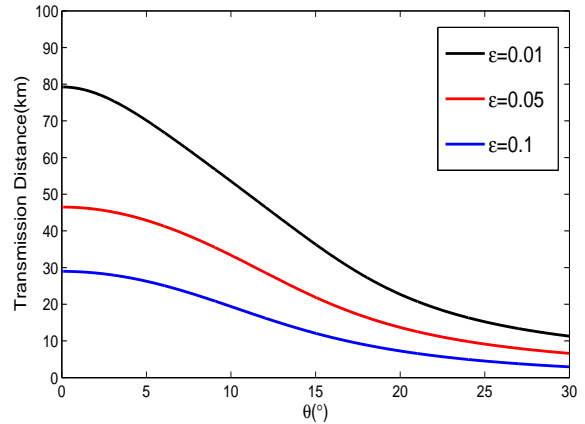}
\caption{The maximum transmittance distance versus measurement angular error. From left to right, the curves correspond to
$\varepsilon  = 0.1$, $\varepsilon  = 0.05$ and $\varepsilon  = 0.01$. The other parameters of systems: modulation variance is $V=20$, reconciliation efficiency is $\beta  = 0.95$. }\label{fig_5}
\end{figure}

In Fig. \ref{fig_4}, we simulate three curves with three measurement angular errors. Not satisfied with the only effect on the secret key rate, the maximum transmittance distance is also computed within a broader range of measurement angular errors.
Thus, we compute the maximum transmittance distance as a function of measurement angular error, and the simulation results are displayed in Fig. \ref{fig_5}.
Different curves represent the results under the influence of different excess noise. For each curve, the maximum transmittance distance decreases significantly. We show the measurement angular error of $30^\circ$ for a complete display of measurement angular errors. However, in practical implementations, we only need to consider measurement angular error as a real small value that is easily ignored. Although the angular error is relatively small, the slope of the curve changes rapidly, indicating that even a tiny measurement angular error would cause a significant loss in transmittance distance in the practical experiment.

\begin{figure}[b]
\centering
\includegraphics[width=8cm,height=6cm]{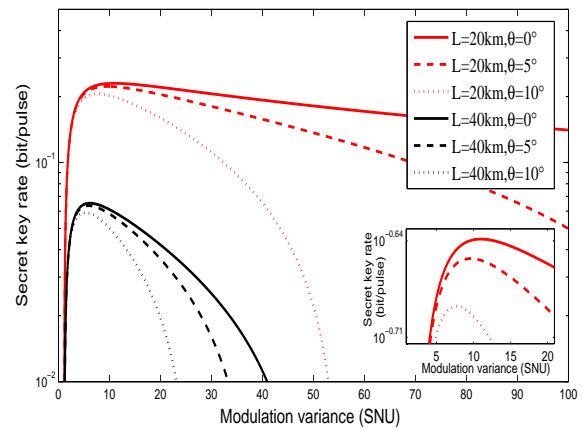}
\caption{The optimal modulation variance for the systems with different measurement angular error. The red and black curves correspond to cases with different distances. Three types of lines, solid, dashed and dotted lines correspond to different measurement angular error.The other parameters of systems: modulation variance is $V=20$, reconciliation efficiency is $\beta  = 0.95$.}\label{fig_6}
\end{figure}

The existence of measurement angular error can also cause the changes of optimal modulation variance; thus, the curves of the secret key rate versus optimal modulation variance with measurement angular error $\theta  = {0^ \circ }$, $\theta  = {5^ \circ }$ and $\theta  = {10^ \circ }$ is simulated, and the simulation result is depicted in Figure \ref{fig_6}. The main figure describes the modulation variance limited within 100, while the inset describes the modulation variance limited within 20 to show the details clearly. It can be intuitively observed that optimal modulation variance decrease with the increase of measurement angular error in different distances.

To sum up, the existence of measurement angular error dramatically affects the performance of the practical CV-QKD system using heterodyne detection.
The existence of measurement angular error will not only reduce the secret key rate of the system but also reduce the transmittance distance; what is worse, the existence of measurement angular error will change the optimal modulation variance of the system, and also reduce the performance of the system,
which further reveals the importance of compensation of measurement angular error.

\section{\label{sec:V}Conclusion}

This paper analyses the effect of measurement angular error caused by the imperfect phase shifter in CV-QKD using heterodyne detection.
The imperfection of the phase shifter is recognized from the detection results of the practical experimental implementation, and the corresponding EB model is proposed based on the PM model.
In order to avoid the potential attack caused by the imperfect phase shifter, we propose a method to estimate the measurement angular error, which can then be compensated through data processing.
Practical security analysis shows channel parameters will be misestimated due to such an error, which further causes a drastic decrease in the achievable secret key rate.
Numerous simulation is provided including the measurement angular error as well as the case after the compensation. Simulation results suggest that the performance of the practical systems can be significantly improved when properly dealing with the measurement angular error.
It should be noted that our analysis can be directly extended to support other CV-QKD systems that rely on the heterodyne detection schemes, for instance, in  LLO-CV-QKD \cite{huang2015high,huang2015continuous}.
Undoubtedly, it is worth observing that our work is to strengthen practical security resulted from devices' imperfection.

\begin{acknowledgments}
This work was supported by the Key Program of National Natural Science Foundation of China under Grant No. 61531003, National Natural Science Foundation of China under Grant No. 62001041, China Postdoctoral Science Foundation under Grant No. 2020TQ0016, and the Fund of State Key Laboratory of Information Photonics and Optical Communications.
\end{acknowledgments}

\appendix

\section{\label{sec:VI}Calculation of the secret key rate}

The secret key rate is conducted under the collective attack for simplicity, we only show the secret key rate for reverse reconciliation
 \cite{grosshans2003quantum}.
\begin{equation}\label{eq.21}
K = \beta {I_{AB}} - {\chi _{BE}},
\end{equation}
where $\beta$ is the reconciliation efficiency, $I_{AB}$ is the mutual information between Alice and Bob, ${\chi _{BE}}$ is the maximum information available to Eve on Bob's key which is bounded by Holevo quantity. The covariance matrix of Alice and the mode that just comes out of the channel ${\gamma _{AB}}$ is
\begin{align}\label{eq.22}
{\gamma _{AB}} &= \left( {\begin{array}{*{20}{c}}
{V \cdot I}&{\sqrt {T \left( {{V^2} - 1} \right)}  \cdot {\sigma _z}}\notag\\
{\sqrt {T \left( {{V^2} - 1} \right)}  \cdot {\sigma _z}}&{\left( {T V + \left( {1 - T} \right)\omega} \right) \cdot I}
\end{array}} \right)\\
: &= \left( {\begin{array}{*{20}{c}}
  {a \cdot I}&{c \cdot {\sigma _z}} \\
  {c \cdot {\sigma _z}}&{b \cdot I}
\end{array}} \right).
\end{align}
where $I = diag(1,1)$, ${\sigma _z} = diag(1, - 1)$, and Eve's variance $\omega$
can be described by
$\omega  = T\varepsilon /(1 - T) + 1$. And we define
$a = V$, $b = TV + \left( {1 - T} \right)\omega $,
$c = \sqrt {T\left( {{V^2} - 1} \right)} $ for simplicity.

Then mode B is divided with mode $B_1$ and $B_2$ by BS. Such a transformation can be expressed as
\begin{equation}\label{eq.23}
{\gamma _{A{B_1}{B_2}}} = {Y_{BS}} \cdot \left( {{\gamma _{AB}} \oplus I} \right) \cdot {Y_{BS}}^T,
\end{equation}
where the matrix ${Y_{BS}}$ describes the BS transformation that acts on only the mode $B$. It is written as
\begin{gather}\label{eq.24}
{Y_{BS}} = I \oplus {S_{BS}},
\end{gather}
where
${S_{BS}} = \left( {\begin{array}{*{20}{c}}
  {\frac{1}{{\sqrt 2 }}I}&{\frac{1}{{\sqrt 2 }}I} \\
  { - \frac{1}{{\sqrt 2 }}I}&{\frac{1}{{\sqrt 2 }}I}
\end{array}} \right)$.
In the same way, the covariance matrix
${\gamma _{A{B_1}{B_3}}}$ after the phase shifter operation becomes
\begin{equation}
\label{eq.25}{\gamma _{A{B_1}{B_3}}} = Y_{PS} \cdot {\gamma _{A{B_1}{B_2}}} \cdot Y{_{PS}^T}, \\
\end{equation}
where ${Y_{PS}} = {I_4} \oplus {S_{PS}}$.

In heterodyne-detection scenario, both data X and P are used to extract keys, thus the mutual information between Alice
and Bob should be ${I_{AB}} = I_{AB}^x + I_{AB}^p$, where
\begin{equation}\label{eq.27}
\left\{ {\begin{array}{*{20}{c}}
{I_{AB}^x = \frac{1}{2}{{\log }_2}\left[ {\frac{{V_A^x}}{{V_{A|B}^x}}} \right]},\\
{I_{AB}^p = \frac{1}{2}{{\log }_2}\left[ {\frac{{V_A^p}}{{V_{A|B}^p}}} \right]},
\end{array}} \right.
\end{equation}
where $V_A^{x(p)} = \frac{1}{2}(a + 1)$, and $V_{A|B}^{x(p)}$ can be calculated from
\begin{equation}\label{eq.28}
V_{A|B}^{x(p)} = \left\langle {{A_{x(p)}}^2} \right\rangle  - \frac{{{\left\langle {{A_{x(p)}}{B_{x(p)}}} \right\rangle }}^2}{{\left\langle {{B_{x(p)}}^2} \right\rangle }}.
\end{equation}

In the following part, we provide the calculation for
${\chi _{BE}}$
\begin{equation}\label{eq.29}
{\chi _{BE}} = S\left( E \right) - S\left( {E|B} \right).
\end{equation}

Here we assume that Eve can purify the whole system, so we can get
$S\left( E \right) = S\left( {AB} \right)$, which is exactly the same as the standard CV-QKD GG02 protocol \cite{grosshans2003quantum}. The symplectic eigenvalues are given by
\begin{equation}\label{eq.30}
{\lambda _{1,2}} = \sqrt {\frac{1}{2}\left( {A \pm \sqrt {{A^2} - 4{B^2}} } \right)},
\end{equation}
where
$A = {a^2} + {b^2} - 2{c^2},B = ab - {c^2}$.

We also assume that after Bob performs projective measurements on his modes, the remaining system $AE$ is pure,
so we get
$S\left( {E|{B_1}{B_3}} \right) = S\left( {A|{B_1}{B_3}} \right)$. $S\left( {E|{B_1}{B_3}} \right)$ is calculated in two steps by first performing P-measurement on
mode $B_3$ and then performing X-measurement on mode $B_1$. The conditional covariance matrix after measuring mode $B_3$ is written as
\begin{equation}\label{eq.31}
{\gamma _{A{B_1}|B_3^p}} = {\gamma _{A{B_1}}} - {C_{A{B_1}{B_3}}}{\left( {{X_p}{\gamma _{{B_3}}}{X_p}} \right)^{MP}}C_{A{B_1}{B_3}}^T,
\end{equation}
where ${C_{A{B_1}{B_3}}}$ and ${\gamma _{A{B_1}}}$ are the submatrices of the covariance matrix
${\gamma _{A{B_1}{B_3}}}$,
${X_p} = diag(0,1)$, $MP$ denotes the Moore–Penrose inverse of a matrix.

Then the X-measurement is performed on mode $B_1$, with the post-measurement covariance matrix
${\gamma _{A|B_3^pB_1^x}}$, which reads
\begin{equation}\label{eq.32}
{\gamma _{A|B_3^pB_1^x}} = {\gamma _{A{B_1}|B_3^p}} - {C_{A{B_1}}}{\left( {{X_x}{\gamma _{{B_1}}}{X_x}} \right)^{MP}}C_{A{B_1}}^T,
\end{equation}
where ${C_{A{B_1}}}$ and ${\gamma _{{B_1}}}$ can be acquired by Eq. (\ref{eq.31}), and ${X_x} = diag(1,0)$. And the symplectic eigenvalue can be calculated by Eq. (\ref{eq.32}), the expressions for the entropies ${\chi _{BE}}$ can be further simplified as follows
\begin{equation}\label{eq.33}
{\chi _{BE}} = \sum\limits_{i = 1}^2 {G\left( {\frac{{{\lambda _i} - 1}}{2}} \right)}  - G\left( {\frac{{{\lambda _3} - 1}}{2}} \right),
\end{equation}
where
$G\left( x \right) = \left( {x + 1} \right){\log _2}\left( {x + 1} \right) - x{\log _2}x$, ${\lambda _{1,2}}$ and ${\lambda _3}$ are the symplectic eigenvalues of ${\gamma _{AB}}$ and $\gamma _{A|B_3^pB_1^x}$ respectively.


\bibliographystyle{plain}
\bibliography{apssamp}

\providecommand{\noopsort}[1]{}\providecommand{\singleletter}[1]{#1}%
\begin{thebibliography}{10}

\bibitem{chu2021practical}
Binjie Chu, Yichen Zhang, Yundi Huang, Song Yu, Ziyang Chen, and Hong Guo.
\newblock Practical source monitoring for continuous-variable quantum key
  distribution.
\newblock {\em Quantum Science and Technology}, 6(2):025012, 2021.

\bibitem{corvaja2017phase}
Roberto Corvaja.
\newblock Phase-noise limitations in continuous-variable quantum key
  distribution with homodyne detection.
\newblock {\em Physical Review A}, 95(2):022315, 2017.

\bibitem{furrer2014reverse}
Fabian Furrer.
\newblock Reverse-reconciliation continuous-variable quantum key distribution
  based on the uncertainty principle.
\newblock {\em Physical Review A}, 90(4):042325, 2014.

\bibitem{gehring2015implementation}
Tobias Gehring, Vitus H{\"a}ndchen, J{\"o}rg Duhme, Fabian Furrer, Torsten
  Franz, Christoph Pacher, Reinhard~F Werner, and Roman Schnabel.
\newblock Implementation of continuous-variable quantum key distribution with
  composable and one-sided-device-independent security against coherent
  attacks.
\newblock {\em Nature communications}, 6(1):1--7, 2015.

\bibitem{grosshans2003quantum}
Fr{\'e}d{\'e}ric Grosshans, Gilles Van~Assche, J{\'e}r{\^o}me Wenger, Rosa
  Brouri, Nicolas~J Cerf, and Philippe Grangier.
\newblock Quantum key distribution using gaussian-modulated coherent states.
\newblock {\em Nature}, 421(6920):238--241, 2003.

\bibitem{guo2021toward}
Hong Guo, Zhengyu Li, Song Yu, and Yichen Zhang.
\newblock Toward practical quantum key distribution using telecom components.
\newblock {\em Fundamental Research}, 1(1):96--98, 2021.

\bibitem{huang2016field}
Duan Huang, Peng Huang, Huasheng Li, Tao Wang, Yingming Zhou, and Guihua Zeng.
\newblock Field demonstration of a continuous-variable quantum key distribution
  network.
\newblock {\em Optics letters}, 41(15):3511--3514, 2016.

\bibitem{huang2015high}
Duan Huang, Peng Huang, Dakai Lin, Chao Wang, and Guihua Zeng.
\newblock High-speed continuous-variable quantum key distribution without
  sending a local oscillator.
\newblock {\em Optics letters}, 40(16):3695--3698, 2015.

\bibitem{huang2016long}
Duan Huang, Peng Huang, Dakai Lin, and Guihua Zeng.
\newblock Long-distance continuous-variable quantum key distribution by
  controlling excess noise.
\newblock {\em Scientific reports}, 6(1):1--9, 2016.

\bibitem{huang2015continuous}
Duan Huang, Dakai Lin, Chao Wang, Weiqi Liu, Shuanghong Fang, Jinye Peng, Peng
  Huang, and Guihua Zeng.
\newblock Continuous-variable quantum key distribution with 1 mbps secure key
  rate.
\newblock {\em Optics express}, 23(13):17511--17519, 2015.

\bibitem{huang2014quantum}
Jing-Zheng Huang, S{\'e}bastien Kunz-Jacques, Paul Jouguet, Christian
  Weedbrook, Zhen-Qiang Yin, Shuang Wang, Wei Chen, Guang-Can Guo, and Zheng-Fu
  Han.
\newblock Quantum hacking on quantum key distribution using homodyne detection.
\newblock {\em Physical Review A}, 89(3):032304, 2014.

\bibitem{huang2013quantum}
Jing-Zheng Huang, Christian Weedbrook, Zhen-Qiang Yin, Shuang Wang, Hong-Wei
  Li, Wei Chen, Guang-Can Guo, and Zheng-Fu Han.
\newblock Quantum hacking of a continuous-variable quantum-key-distribution
  system using a wavelength attack.
\newblock {\em Physical Review A}, 87(6):062329, 2013.

\bibitem{jouguet2013preventing}
Paul Jouguet, S{\'e}bastien Kunz-Jacques, and Eleni Diamanti.
\newblock Preventing calibration attacks on the local oscillator in
  continuous-variable quantum key distribution.
\newblock {\em Physical Review A}, 87(6):062313, 2013.

\bibitem{jouguet2013experimental}
Paul Jouguet, S{\'e}bastien Kunz-Jacques, Anthony Leverrier, Philippe Grangier,
  and Eleni Diamanti.
\newblock Experimental demonstration of long-distance continuous-variable
  quantum key distribution.
\newblock {\em Nature photonics}, 7(5):378--381, 2013.

\bibitem{leverrier2015composable}
Anthony Leverrier.
\newblock Composable security proof for continuous-variable quantum key
  distribution with coherent states.
\newblock {\em Physical review letters}, 114(7):070501, 2015.

\bibitem{leverrier2017security}
Anthony Leverrier.
\newblock Security of continuous-variable quantum key distribution via a
  gaussian de finetti reduction.
\newblock {\em Physical review letters}, 118(20):200501, 2017.

\bibitem{leverrier2013security}
Anthony Leverrier, Ra{\'u}l Garc{\'\i}a-Patr{\'o}n, Renato Renner, and
  Nicolas~J Cerf.
\newblock Security of continuous-variable quantum key distribution against
  general attacks.
\newblock {\em Physical review letters}, 110(3):030502, 2013.

\bibitem{liu2020imperfect}
Weiqi Liu, Jinye Peng, Jin Qi, Zhengwen Cao, and Chen He.
\newblock Imperfect basis choice in continuous-variable quantum key
  distribution.
\newblock {\em Laser Physics Letters}, 17(5):055203, 2020.

\bibitem{liu2017imperfect}
Wenyuan Liu, Xuyang Wang, Ning Wang, Shanna Du, and Yongmin Li.
\newblock Imperfect state preparation in continuous-variable quantum key
  distribution.
\newblock {\em Physical Review A}, 96(4):042312, 2017.

\bibitem{ma2013local}
Xiang-Chun Ma, Shi-Hai Sun, Mu-Sheng Jiang, and Lin-Mei Liang.
\newblock Local oscillator fluctuation opens a loophole for eve in practical
  continuous-variable quantum-key-distribution systems.
\newblock {\em Physical Review A}, 88(2):022339, 2013.

\bibitem{painchaud2009performance}
Yves Painchaud, Michel Poulin, Michel Morin, and Michel T{\^e}tu.
\newblock Performance of balanced detection in a coherent receiver.
\newblock {\em Optics express}, 17(5):3659--3672, 2009.

\bibitem{pirandola2015high}
Stefano Pirandola, Carlo Ottaviani, Gaetana Spedalieri, Christian Weedbrook,
  Samuel~L Braunstein, Seth Lloyd, Tobias Gehring, Christian~S Jacobsen, and
  Ulrik~L Andersen.
\newblock High-rate measurement-device-independent quantum cryptography.
\newblock {\em Nature Photonics}, 9(6):397--402, 2015.

\bibitem{qin2016quantum}
Hao Qin, Rupesh Kumar, and Romain All{\'e}aume.
\newblock Quantum hacking: Saturation attack on practical continuous-variable
  quantum key distribution.
\newblock {\em Physical Review A}, 94(1):012325, 2016.

\bibitem{walk2016experimental}
Nathan Walk, Sara Hosseini, Jiao Geng, Oliver Thearle, Jing~Yan Haw, Seiji
  Armstrong, Syed~M Assad, Jiri Janousek, Timothy~C Ralph, Thomas Symul, et~al.
\newblock Experimental demonstration of gaussian protocols for one-sided
  device-independent quantum key distribution.
\newblock {\em Optica}, 3(6):634--642, 2016.

\bibitem{wang2020high}
Heng Wang, Yaodi Pi, Wei Huang, Yang Li, Yun Shao, Jie Yang, Jinlu Liu, Chenlin
  Zhang, Yichen Zhang, and Bingjie Xu.
\newblock High-speed gaussian-modulated continuous-variable quantum key
  distribution with a local local oscillator based on pilot-tone-assisted phase
  compensation.
\newblock {\em Optics Express}, 28(22):32882--32893, 2020.

\bibitem{wang2017efficient}
Xiangyu Wang, Yichen Zhang, Song Yu, Bingjie Xu, Zhengyu Li, and Hong Guo.
\newblock Efficient rate-adaptive reconciliation for continuous-variable
  quantum key distribution.
\newblock {\em Quantum Information \& Computation}, 17(13-14):1123--1134, 2017.

\bibitem{wang2017experimental}
Xuyang Wang, Wenyuan Liu, Pu~Wang, and Yongmin Li.
\newblock Experimental study on all-fiber-based unidimensional
  continuous-variable quantum key distribution.
\newblock {\em Physical Review A}, 95(6):062330, 2017.

\bibitem{weedbrook2012gaussian}
Christian Weedbrook, Stefano Pirandola, Ra{\'u}l Garc{\'\i}a-Patr{\'o}n,
  Nicolas~J Cerf, Timothy~C Ralph, Jeffrey~H Shapiro, and Seth Lloyd.
\newblock Gaussian quantum information.
\newblock {\em Reviews of Modern Physics}, 84(2):621, 2012.

\bibitem{zhang2012towards}
Bo~Zhang, Christian Malouin, and Theodore~J Schmidt.
\newblock Towards full band colorless reception with coherent balanced
  receivers.
\newblock {\em Optics Express}, 20(9):10339--10352, 2012.

\bibitem{zhang2020long}
Yichen Zhang, Ziyang Chen, Stefano Pirandola, Xiangyu Wang, Chao Zhou, Binjie
  Chu, Yijia Zhao, Bingjie Xu, Song Yu, and Hong Guo.
\newblock Long-distance continuous-variable quantum key distribution over
  202.81 km of fiber.
\newblock {\em Physical review letters}, 125(1):010502, 2020.

\bibitem{zhang2019continuous}
Yichen Zhang, Zhengyu Li, Ziyang Chen, Christian Weedbrook, Yijia Zhao, Xiangyu
  Wang, Yundi Huang, Chunchao Xu, Xiaoxiong Zhang, Zhenya Wang, et~al.
\newblock Continuous-variable qkd over 50 km commercial fiber.
\newblock {\em Quantum Science and Technology}, 4(3):035006, 2019.

\bibitem{zhao2018polarization}
Yijia Zhao, Yichen Zhang, Yundi Huang, Bingjie Xu, Song Yu, and Hong Guo.
\newblock Polarization attack on continuous-variable quantum key distribution.
\newblock {\em Journal of Physics B: Atomic, Molecular and Optical Physics},
  52(1):015501, 2018.

\end{thebibliography}

\end{document}